# Ordered magnetic and quadrupolar states under hydrostatic pressure in orthorhombic $PrCu_2$

T. Naka,[a, *] L. A. Ponomarenko,[b] A. de Visser,[b] A. Matsushita,[a] R. Settai,[c] and Y. Ōnuki[c]

[a]*Materials Engineering Laboratory, National Institute of Materials Science, Tsukuba, Ibaraki 305-0047, Japan*

[b]*Van der Waals-Zeeman Institute, University of Amsterdam, Valckenierstraat 65, 1018 XE, Amsterdam, The Netherlands*

[c]*Faculty of Science, Osaka University, Toyonaka, Osaka 560-0043, Japan*

*Abstract*

We report magnetic susceptibility and electrical resistivity measurements on single-crystalline $PrCu_2$ under hydrostatic pressure, up to 2 GPa, which pressure range covers the pressure-induced Van Vleck paramagnet-to-antiferromagnet transition at 1.2 GPa. The measured anisotropy in the susceptibility shows that in the pressure-induced magnetic state the ordered 4*f*-moments lie in the *ac*-plane. We propose that remarkable pressure effects on the susceptibility and resistivity are due to changes in the quadrupolar state of $O_2^2$ and/or $O_2^0$ under pressure. We present a simple analysis in terms of the singlet-singlet model.

* Corresponding author. Tel.: +81-29-859-2730; fax: +81-29-859-2701; e-mail: NAKA.Takashi@nims.go.jp.

## 1. Introduction

Using the high pressure technique one may tune the volume and the corresponding properties of solids, thereby inducing transformations from one solid state to another, such as non-magnetic-magnetic phase transitions and crystallographic phase transitions. In rare earth compounds, the compression of the crystal modifies the electronic state through one-ion crystal-field and magneto-elastic interactions and/or two-ion interactions, such as quadrupolar coupling and the Ruderman-Kittel-Kasuya-Yoshida (RKKY) interaction mediated by conduction band electrons.[1] The orthorhombic compound $PrCu_2$ ($CeCu_2$-type of crystal structure) may be considered to be an exemplary compound in this respect. $PrCu_2$ has a singlet ground state and undergoes a cooperative Jahn-Teller (JT) transition at $T_{JT}$=7.6 K. The crystal field of local symmetry $C_{2v}$, splits the full $^3H_4$ multiplet of $Pr^{3+}$ into nine singlets. In fact, the magnetic susceptibility approaches a constant value at low temperature. As will be mentioned below, a huge anisotropic magnetostriction [2] accompanies the metamagnetic transition, which occurs for the magnetic field applied parallel to the *ac*-plane.[3, 4] Consequently, considerable pressure effects on the magnetic state are expected in $PrCu_2$. In view of this, the effects of pressure on the magnetic susceptibility, specific heat, electric resistivity and lattice constants were investigated recently.[5] A pseudo-uniaxial compression, with compressibilities $\kappa_a \cong \kappa_c < \kappa_b$, was observed, but no signal of a structural transition up to 4 GPa. However, at $P$>1.2 GPa the susceptibility showed a remarkable maximum at $T=T_{max}$. At low temperature, an upturn in the specific heat develops with pressure, which possibly can be attributed to the emergence of a spontaneous magnetic field. The estimated spontaneous fields are comparable to those in other magnetically-ordered 4f electron systems.[5] Around a pressure of $P$=1.2 GPa the resistivity at low temperatures is enhanced. These findings suggest that pressure-induced magnetism appears at $P$>1.2 GPa. In Fig. 1 we show the pressure-temperature ($P$-$T$) phase diagram of $PrCu_2$. The magnetic transition temperature (or $T_{max}$) increases steeply near 1.2 GPa and increases monotonously at higher pressure. At $P$>1.7 GPa $T_{max}$ becomes larger than $T_{JT}$, the latter being nearly pressure independent but exhibiting a weak maximum at $P$=1.2 GPa.[5]

This magnetic behaviour reminds one of the pressure-induced antiferromagnetism appearing at $P_c$=3.0 GPa [6] in the cubic material PrSb, which is simply characterized by “induced-moment magnetism” (IMM). For the singlet ground state no magnetic order develops when the interatomic exchange interaction, $K$, remains below a certain critical threshold, $K_c$. For $K > K_c$ the so-called induced-moment transition occurs at finite temperature. However, when strong hyperfine interactions between nuclear and electronic moments exist, even for $K < K_c$, both the moments can undergo a magnetic ordering at lower temperature.[1] In $PrCu_2$ at ambient pressure, actually, hyperfine coupled electron-nuclear magnetism (HCENM) appears at very low temperatures (i.e. below 54 mK).[1, 7] Kawarazaki *et al.*[7] resolved the magnetic structure by neutron scattering. The electronic and nuclear spins on the Pr-ion are sinusoidally modulated in magnitude with a modulation vector $\boldsymbol{Q}_\mathbf{m}$=(0.24, 0, 0.68) and point almost along the *a*-axis. Consequently, it has been speculated that in $PrCu_2$ the phase transition at $P$=1.2 GPa is of the HCENM-to-IMM.[5, 8]

In the past decade, it has become clear that quadrupolar interactions play an important role in the magnetism in $PrCu_2$, especially, since the metamagnetic transition for magnetic fields in the *ac*-plane is accompanied by very large magnetostrictions.[2-4] In addition to the crystal field interaction, the quadrupole-quadrupole couplings of $O_{xy}$ and $O_2^2$, are essential for the reproduction of the features, such as, the co-operative JT- and the metamagnetic transitions, respectively.[4] The former is characterized to be a ferroquadrupolar ordering of $O_{xy}$. On the other hand, the rotation of the quadrupolar moment of $O_2^2$ under magnetic field is responsible for the latter. Unexpectedly, in $PrCu_2$ at $T$=65 K magnetic ordering was observed by a comprehensive muon-spin-rotation/relaxation (μSR) study.[9] Spontaneous internal magnetic fields appear below 65 K. It was speculated that the magnetic structure is similar to that of the low-temperature HCENM phase at $T$<54 mK. The estimated ordered electronic moment of 0.29 $\mu_B$ is smaller than that of 0.54 $\mu_B$ quoted by Kawarazaki *et al.* It was suggested that a nonmagnetic interaction related to a collective state of $O_{xy}$ attributes to the magnetic ordering at $T$<65 K.

The magnetic state in the pressure-induced phase at $P$>1.2 GPa is not understood yet. In this work, we report a high-pressure study of the magnetic susceptibilities along the different crystal axes in order to characterize the magnetic structure at high pressure. Additionally, the electrical resistivity, $\rho(T)$, is investigated down to 0.4 K in order to confirm the pressure-temperature ($P$-$T$) phase diagram and to determine the pressure effect on the crystalline electric field splitting of the low-lying crystal field states. We found that even in the paramagnetic region the anomaly of the $\chi$-$P$ curve around $P_c$=1.2 GPa survives. This anomaly preceding the transition to the magnetic state is possibly related to the quadrupolar state at high pressure.

## 2. Experimental methods

Single crystalline samples were made by a Czochralski pulling method in an induction furnace under helium gas atmosphere.[3] The magnetization was measured at high pressure up to $P$=1.8 GPa using the Faraday method in the temperature range 2-80 K. Measurements were done in a superconducting magnet system, which consists of a main solenoid coil producing a main field, $\boldsymbol{H}_{main}$, and a set of gradient coils located at the top and bottom sides of the main coil producing a field gradient, $dH_z/dz$, along the vertical axis (the $z$-axis) at the sample position. While supplying a current to the magnet coils, the force being proportional to $\chi(dH_z/dz)H$ can be detected by the force balance. In contrast to the resistive magnet system, in which the magnetization $\boldsymbol{M}$ is perpendicular to the force, $\boldsymbol{F}_z$, the superconducting magnet system produces fields, $\boldsymbol{H}$ and $\boldsymbol{M}$ parallel to $\boldsymbol{F}_z$. In this configuration it is possible to measure the anisotropy in the magnetization, since it is difficult for the easy axis of magnetization to rotate towards the direction of the magnetic field when the magnetization along the hard axis is measured.

The single crystalline sample was put into a Teflon cell with Fluorinert as pressure transmitting media. It was then pressurized in a piston-cylinder-type clamp made from a CuBe alloy. To obtain $\chi(T)$ in various magnetic fields and $T_{max}$ and $T_{JT}$, we took data at $H$= 4, 10 and 20 kOe. The resistivity was measured using a standard ac 4-probe method using a Linear Research resistance bridge (model LR-700). In this case a hybrid clamp cell made from NiCrAl and CuBe alloys, which can generate pressures over 2 GPa, was used. The sample was mounted on a specially designed plug and put inside a Teflon cell with the pressure-transmitting medium Flourinert [5]. The absorption pump operated $^3$He cryogenic system is described elsewhere.[10]

## 3. Results and discussion

### 3-1. Magnetic susceptibility

In Fig. 2 we show the susceptibility under pressure for magnetic fields along the *a*, *b* and *c*-axis. The susceptibility along the *a*-axis, $\chi_a(T)$, shows a pronounced maximum at $T_{\max}$ (Fig. 2(a)) for pressures $P$>1.2 GPa. In contrast, $\chi_b(T)$ shows no anomaly at $T_{\max}$ over the entire pressure range, however, a kink is observed at $T_{\rm JT}$. For $\chi_c(T)$ the pressure-induced anomaly at $T_{\max}$ is less pronounced. The observed anisotropy in the susceptibility suggests that the ordered 4f-moments lie in the *ac*-plane, approximately parallel to the *a*-axis. This is consistent with the observation that at ambient pressure the ordered nuclear and electronic moments are oriented approximately parallel to the *a*-axis and sinusoidally modulated in magnitude below $T_N$=54 mK.[11]

The JT-transition temperature as a function of pressure, $T_{\rm JT}(P)$, can be obtained from the kink point of $\chi_b(T)$ (see inset Fig.2b). $T_{\rm JT}$ depends weakly on pressure and exhibits a weak maximum at 1.2 GPa (Fig. 1). On the other hand, $T_{\max}$ appears at 1.2 GPa and increases rapidly with pressure. Figures 3 (a)-(c) show the pressure dependencies of the susceptibilities along the *a*-, *b*- and *c*-axis, respectively, above and below $T_{\rm JT}$ measured in a field of $H$=4 kOe. $\chi_a(P)$ and $\chi_c(P)$ exhibit pronounced maxima around $P$=1 GPa in the lower temperature range, while $\chi_b(P)$ decreases linearly with pressure over the entire pressure range. This contrasting behaviour between the pressure dependencies in the *ac*-plane and the *b*-axis seems to be in line with that observed in the dynamic phenomena of the quadrupolar moments that lie in the *ac*-plane, as reported in Ref. 2-4. Also, the linear thermal expansion coefficients $\alpha_a$ and $\alpha_c$ show anomalously sharp peaks at $T_{\rm JT}$, while $\alpha_b$ exhibits a weaker anomaly.[2] The metamagnetic behaviour with the anomalously large magnetostriction occurs for the magnetic fields applied parallel to the *ac*-plane.[3] It was confirmed that the easy (*a*-) and the hard (*c*-) axis directions of the magnetization interchange through the metamagnetic transition.[3-4] Based on a theoretical model including quadrupole-quadrupole interactions, it is derived that the cooperative JT transition results from a quadrupolar ordering of $O_{xy}$, while the metamagnetic transition is due to the rotation of the quadrupolar moment of $O_2^2$ under magnetic field [4]. The quadrupolar state of $O_2^2$ is strongly coupled to magnetic moments in the *ac*-plane. The maxima in $\chi_a(P)$ and $\chi_c(P)$ at around $P$=1 GPa are observed not only below but also well above $T_{\max}$, as shown in Figs. 3(a) and (c). Therefore, it is plausible that the quadrupolar state of $O_2^2$ changes at $P$>1.2 GPa.

Notably, the co-operative JT transition, that is, the quadrupolar ordering of $O_{\rm xy}$ does not affect considerably the magnetization in the *ac*-plane, and, hence, the $O_2^2$ state. It is likely that these states behave independently of each other. This situation is preserved also at high pressure. In fact, in contrast to the magnetism in the *ac*-plane, $T_{\rm JT}(P)$ is quite insensitive to pressure, as can be seen in Fig. 1.

### 3-2. Electric resistivity and the singlet-singlet model

In Figure 4 we show $\rho(T)$ at various pressures. Under pressure $\rho$ increases at low temperature. Around $P$=1 GPa a minimum in the resistivity is observed (see inset of Fig. 4). When plotted versus $T^2$ (Fig.5) the low-temperature data clearly show that $\rho(T)$ deviates from the $T^2$-dependence with increasing temperature. Note that in a previous paper we had claimed that the resistivity obeys the characteristics of the Fermi-liquid $\rho(T)=\rho_0+AT^2$ down to 2 K.[5] Our new result indicates that there is a thermally excited component in the resistivity, due to the low lying excitations in the crystal field states of $PrCu_2$.

In order to obtain information about the crystal field splitting in $PrCu_2$ from the $\rho(T)$–data, we fitted the data below 4 K to the following simple equation:

$$\rho(T) = \rho_0 + AT^2 + \sum_i \rho_i \frac{\exp(-\Delta_i / k_B T)}{\sum \exp(-\Delta_i / k_B T)} \quad (1)$$

Here the fitting parameters are the residual resistivity $\rho_0$, the coefficient of the $T^2$ term, $A$, a component of the thermally excited state in the resistivity, $\rho_i$, and the corresponding crystal field splitting $\Delta_i$. The third term yields the contributions from thermally excited singlet states. We assume that $\rho_i$ is independent of temperature. Based on the crystal field level scheme for the four lowest levels as shown in Fig. 6 [12] we have $\rho_1$, $\rho_2 << \rho_3$ at the lowest pressure. This result makes plausibly that the singlet-singlet model with the ground and the third excited states describes the low-temperature resistivity. Otherwise, the first and second excited states respond mainly to the JT transition, as pointed out previously.[13]

Figures 7 (a) and (b) show the fitting parameters as a function of pressure obtained on the basis of the singlet-singlet model. As previously indicated,[5] $\rho_0$ and $A$ are enhanced anomalously at $P_c$ and increase with pressure above $P_c$, while both $\rho_1(P)$ and $\Delta(P)$ seem to have a weak minimum at $P_c$. Remarkably, the extrapolated value of $\Delta$~14K at $P$=0 corresponds well with the energy splitting between the ground and the third excited states, as obtained from inelastic neutron scattering experiments.[12] Kawarazaki et al. measured the anisotropy and the dispersion of the singlet-singlet crystal–field exciton down to 1.8 K in $PrCu_2$. The excitons with $J_x$ polarization and with $J_y$ polarization have quite different dispersions. The strong dispersive $J_x$ transition shows a substantial softening of the excitation energy at $\boldsymbol{Q}_m=0.24\boldsymbol{a}^*+0.68\boldsymbol{c}^*$. This is consistent with the observation that the ordered magnetic moment is parallel to the $a$-axis.[11] As mentioned above, it is plausible that the magnetic structure at $P$=0 is preserved in the pressure-induced phase. Additionally, anisotropic RKKY interactions between the 4f electrons, $K_{a,b,c}$, have been reported.[13] The low-lying crystal-field splitting is not sensitive to the application of hydrostatic pressure in $PrCu_2$, as shown in Fig. 7 (b). Therefore, we suggest that the anisotropic pressure dependence of the susceptibilities, $\chi_{a,b,c}$, for the greater part result from the anisotropy in the exchange interactions $K_{a,b,c}$.

Around $P_c$, $\rho(T)$ obviously exhibits a hump at low temperature as shown in the inset of Fig. 4. Such an increase is also observed at an antiferromagnetic transition due to the opening of an energy gap at the Fermi-surface.[14] In this case, the resistivity, $\rho'(T)$, can be written as:

$$\rho'(T) = \frac{\rho(T)}{1 - gm(T)} \quad (2)$$

where $\rho(T)$ is the resistivity of the normal state and $g$ a truncation factor. $m(T)$ is defined as $M_Q(T)/M_Q(0)$, where $M_Q(T)$ is the staggered moment, and is approximately expressed as $m(T)=(1-(T/T_N)^2)^{1/2}$. In the inset of Fig. 4, the calculated $\rho$ in the vicinity of $P_c$ is shown together with the experimental data. The structure with a minimum in $\rho(T)$ is reproduced. At $P$>1.9 GPa the hump in $\rho(T)$ is observed, which can be reproduced by eq. (2), whereas the $\rho(T)$ data is not shown here. The truncation factor $g$ is obtained to be 0.02-0.03 which is considerably smaller than the values 0.5-0.6 obtained for the spin density wave (SDW) transition in Cr and related compounds.[14] This suggests that the low-lying crystal field states of the 4f electrons are weakly hybridized with the conduction electrons. The anomalous points associated with the pressure-induced transition, $T_{anomaly}$, are shown as a function of pressure in Figs. 1 and 9.

### 3-3. Analysis based on the singlet-singlet model

In this section we investigate to what extent the singlet-singlet model can reproduce the pressure-temperature phase diagram of $PrCu_2$ shown in Fig. 1. Before we calculate the magnetic transition temperature as a function of pressure, we recall a simple model for a Van Vleck paramagnet with a singlet ground state and the first excited singlet state located at an energy $\Delta$, the so-called singlet-singlet model. The interaction between the electronic and nuclear magnetic moments is neglected. In this model, the magnetic phase diagram is determined as a function of a characteristic factor $\eta=4K_0\alpha^2/\Delta$, where $K_0$ and $\alpha$ are the inter-atomic electronic exchange interaction and the off-diagonal matrix element of $J_z$ between the two singlets, $\langle 0|J_z|1\rangle$, respectively. It is derived that magnetic order and a spontaneous moment appear abruptly at the critical points $|\eta|=1$ in an accessible temperature range.[1, 8] It was shown in Ref. 5 that in $PrCu_2$ the characteristic parameter $\eta$ reaches the critical value, $|\eta_c|=1$ at $P$=1.2 GPa . But this previous estimate depended strongly on the initial value of $\eta(0)$. Since the value $\eta(0)$ in $PrCu_2$ is controversial,[7, 11] we assume the condition $|\eta_c|=1$ at $P$=1.2 GPa in the following. In the singlet-singlet model, the susceptibility at $T \ll \Delta/k_B$ is expressed as:

$$\chi(0) = \frac{c}{\Delta}\frac{1}{1-\eta} \tag{3}$$

where $c$ is pressure independent. Consequently, the pressure coefficient of $\chi$ consists of the following two terms

$$\frac{d\ln\chi(0)}{dP} = -\frac{d\ln\Delta}{dP} - \frac{d\ln(1-\eta)}{dP} \tag{4}$$

The first term accounts for the crystal field splitting and the second term is due to the enhancement factor, $1-\eta$. As shown in Fig. 8, the signs of the pressure coefficients change abruptly at $P_c$ and $|d\ln\Delta/dP| \ll |d\ln(1-\eta)/dP|$ at $P \geq P_c$. We conclude, therefore, that the antiferromagnetic exchange coupling along the *a*-axis increases anomalously with pressure at the induced-moment transition. Using the values of $d\ln\chi_a/dP$ just above $T_{max}$ and $d\ln\Delta/dP$, we have estimated numerically the transition temperature $T_{mag}^{cal}(P)$. The results are shown in Fig. 1, which reveals that the calculated $T_{mag}^{cal}(P)$ increases more rapidly with pressure than the experimental value.

### 3-4. Magnetic and quadrupolar phase diagram

We speculated above that at the critical pressure of 1.2 GPa, where $\chi_a(P)$ and $\chi_c(P)$ exhibit a maximum, the quadrupolar state of $O_2^2$ seems to change considerably, while no structural transition is observed.[5] This makes it plausible that the pressure-induced magnetic state is stabilized by the pressure-induced quadrupolar state. Figure 9 shows the magnetic transitions in the $P$-$T$ phase diagram. In this figure, the points where $\chi_a(P)$ exhibits a maximum are connected by the solid line, which is denoted as $P_{max}$ in Fig. 9. Since the cooperative Jahn-Teller transition shows no dominant contribution to the pressure-induced magnetic transition $T_{JT}(P)$ is omitted in the figure. The $P_{max}$-$T$ curve seems to make a special point at ($P_c$, $T_c$)≈(1.2 GPa, 5 K) with the $T_{max}$-$P$ curve. On the other hand, at lower pressure, it is likely that $T_{max}$ ($P$) extends to the point, $T_N$ =54 mK and $P$=0. This is supported by the fact that the magnetic structure at higher pressure is similar to that at $P$=0, as mentioned above.

Finally, it is worth to note that the anisotropic compression under hydrostatic pressure, $\kappa_a \cong \kappa_c < \kappa_b$,[5] corresponds effectively with the symmetry strains, $\varepsilon_u=(2\,\varepsilon_{zz}-\varepsilon_{xx}-\varepsilon_{yy})/\sqrt{3}$ and $\varepsilon_v=\varepsilon_{xx}-\varepsilon_{yy}$, which couple to the quadrupolar moments $O_2^0$ and $O_2^2$, respectively.[15] Especially, the uniaxial tendency of the compression may stabilize or destabilize the component of $O_2^0$ in the quadruple state. As a result, the pressure-induced quadrupolar state at $P$>$P_c$ results in the anisotropic reshaping of the Fermi surface, which is possibly suitable to stabilize a SDW state by nesting. We suggest, therefore, that this SDW scenario can describe not only the pressure-induced magnetism involving the 4f moments, but also the extremely high temperature magnetism appearing at $T$=65 K.[9] For the latter, magnetic ordering involving the spins of the highly anisotropic conduction band electrons indicated in Ref. 13 should be taken into account.

## 4. Summary

We have reported susceptibility and transport data of single-crystalline $PrCu_2$ under pressure. The observed anisotropy in the susceptibility indicates that below the pressure-induced magnetic transition the ordered moments lie in the *ac*-plane. In the investigated temperature range, 2K<$T$<80 K, a magnetic anomaly at $P_c$ is observed for magnetic fields directed in the *ac*-plane, while no anomaly results for magnetic fields along the *b*-axis. Near $P_c$ the resistivity data show an anomaly as function of temperature near 0.8 K and above $T_{JT}$. The singlet-singlet model can only in part account for the observed phase diagram. Our new results indicate that possibly a pressure-induced quadrupolar state appears as a precursor of the pressure-induced magnetism. Further investigation with respect to the pressure-induced magnetic and quadrupolar states should be done on the basis of microscopic point of view.

**Figures and figure captions**

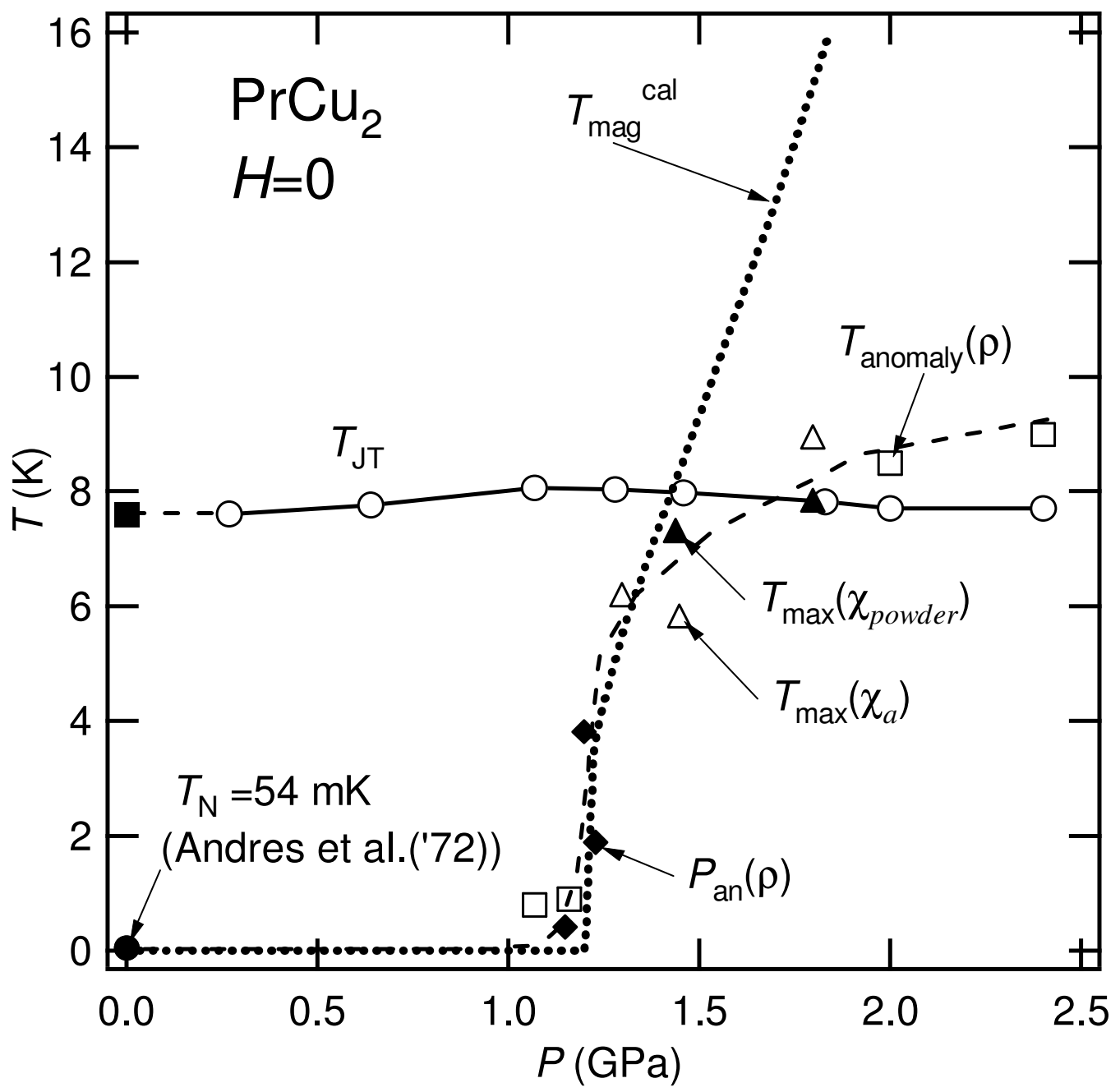

Fig. 1 Pressure-temperature phase diagram of $PrCu_2$ at $H$=0. The dotted line represents $T_{mag}^{cal}(P)$ calculated on basis of the singlet-singlet model. $T_{max}(\chi_a)$, $T_{max}$(powder) and $P_{an}(\rho)$ were obtained previously in Ref. 5. Solid circle and square at $P$=0 were indicated in Refs. 7 and 4, respectively. Dashed and solid lines are guides to the eye.

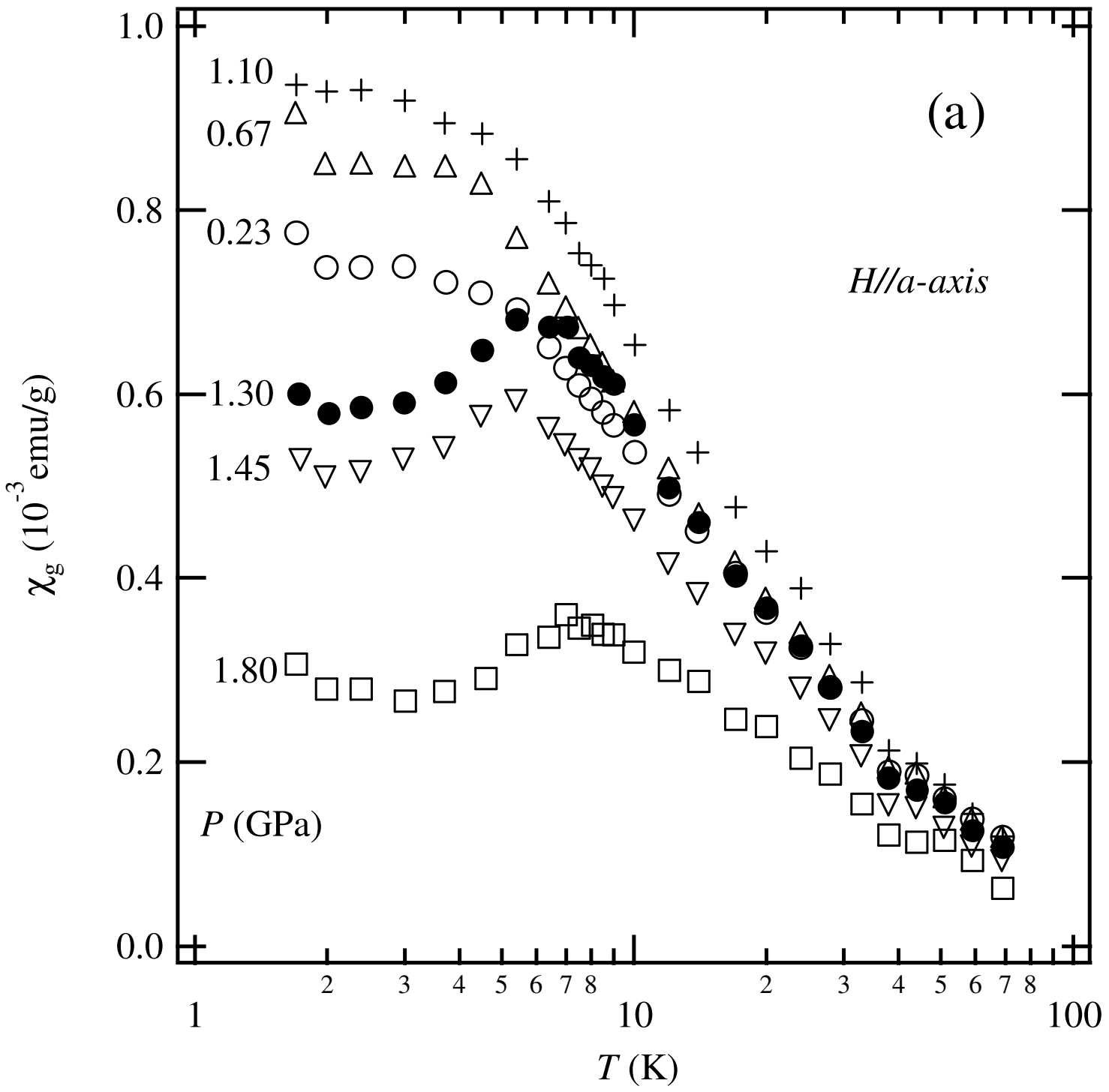

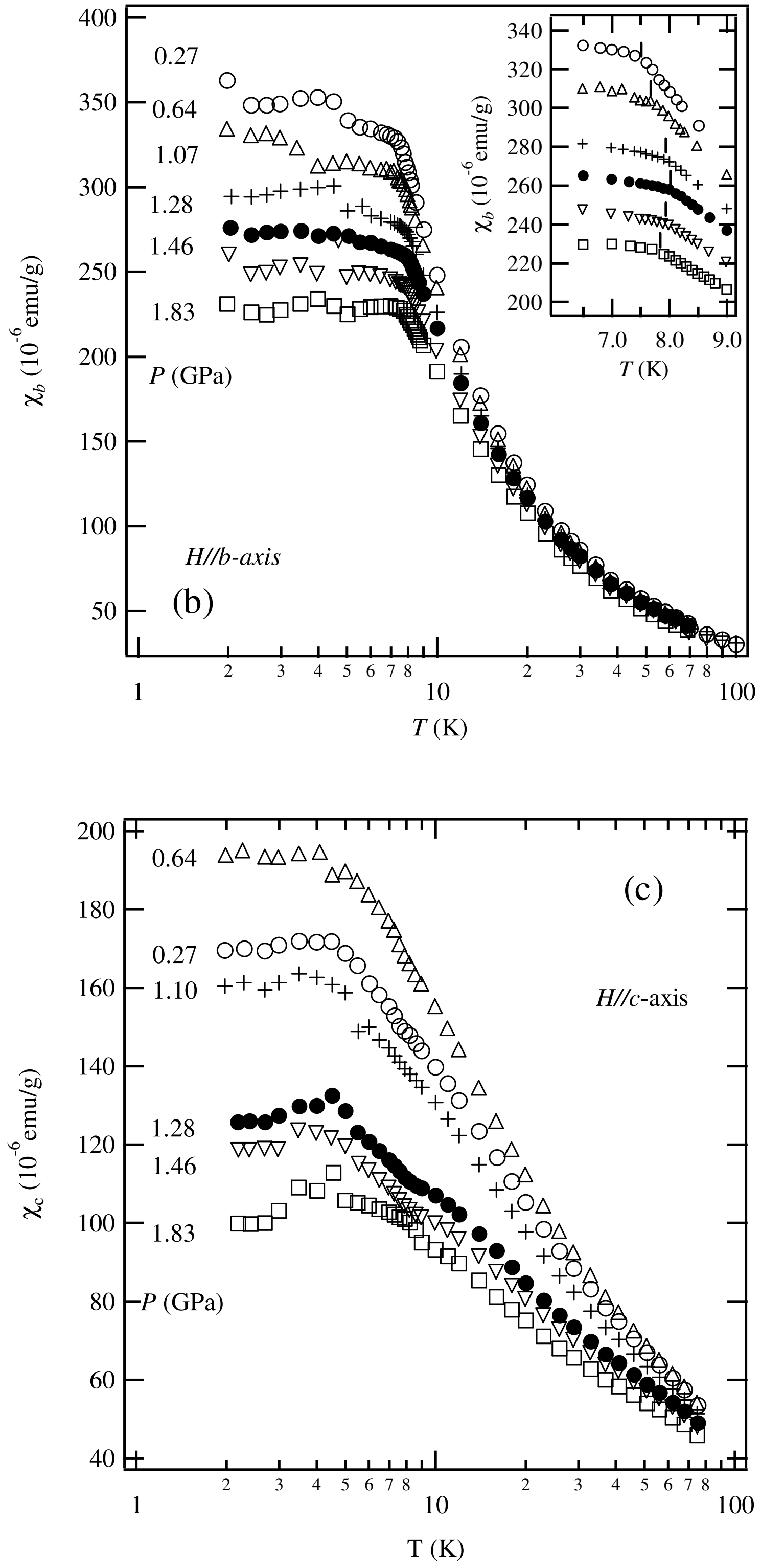

Figs. 2. Magnetic susceptibility of $PrCu_2$ as a function of pressure for applied magnetic fields parallel to the *a*-(a), *b*- (b) and the *c*-axis (c), respectively. The inset of Fig.2(b) shows the Jahn-Teller-transition temperature as a function of pressure, determined by the kink in $\chi_c(T)$.

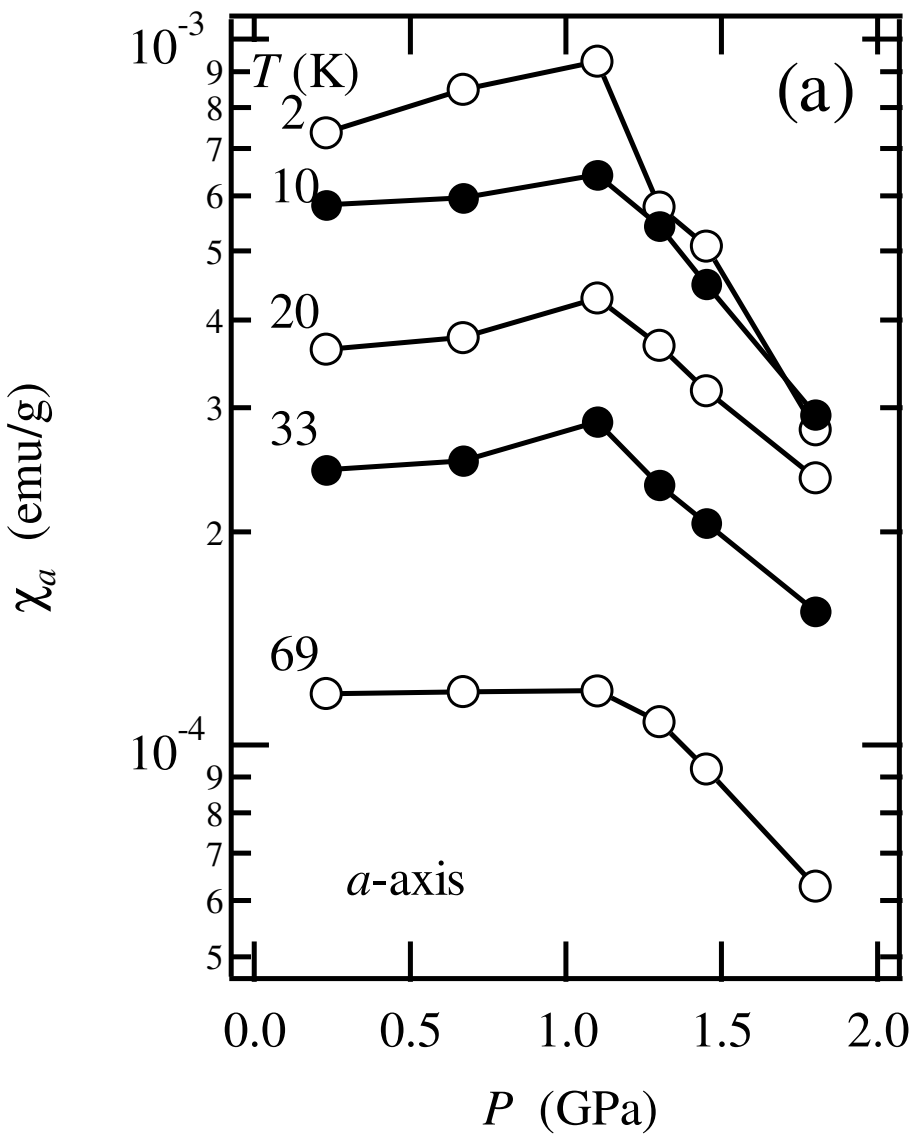

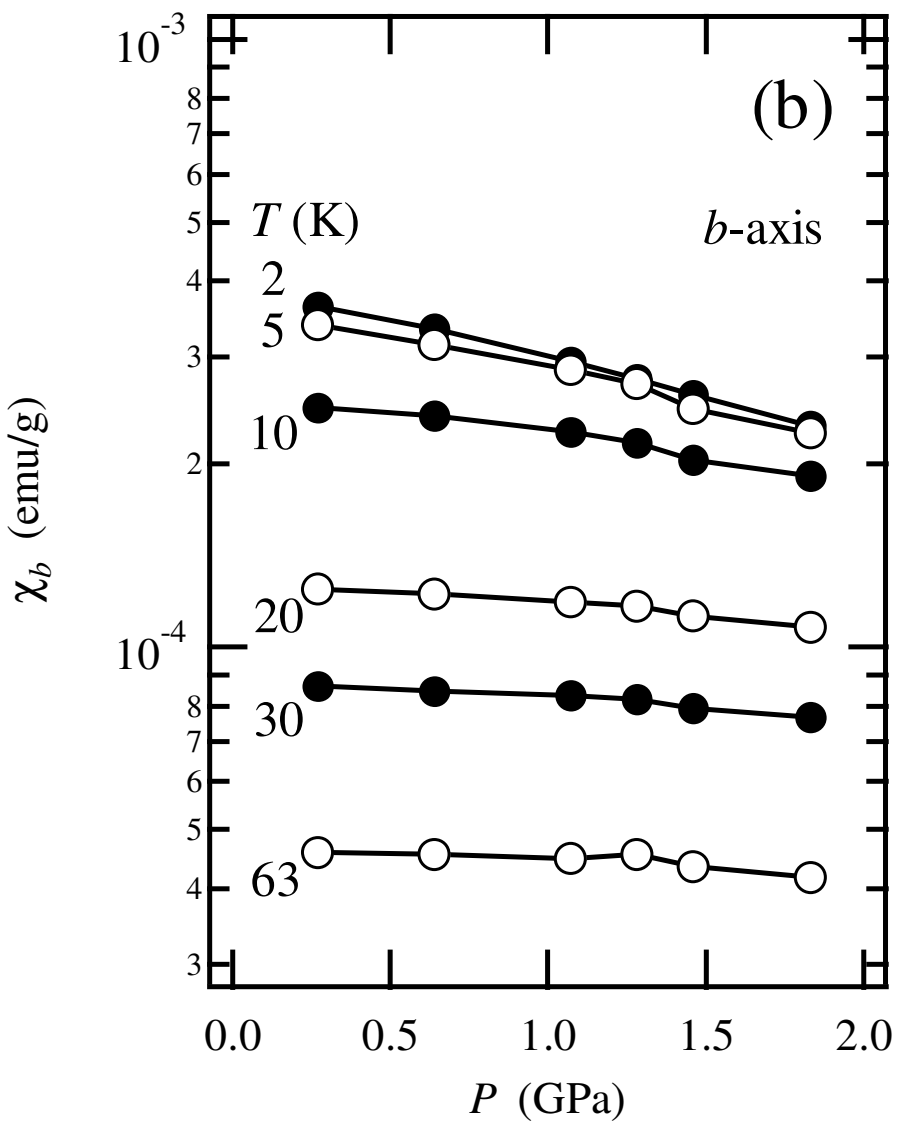

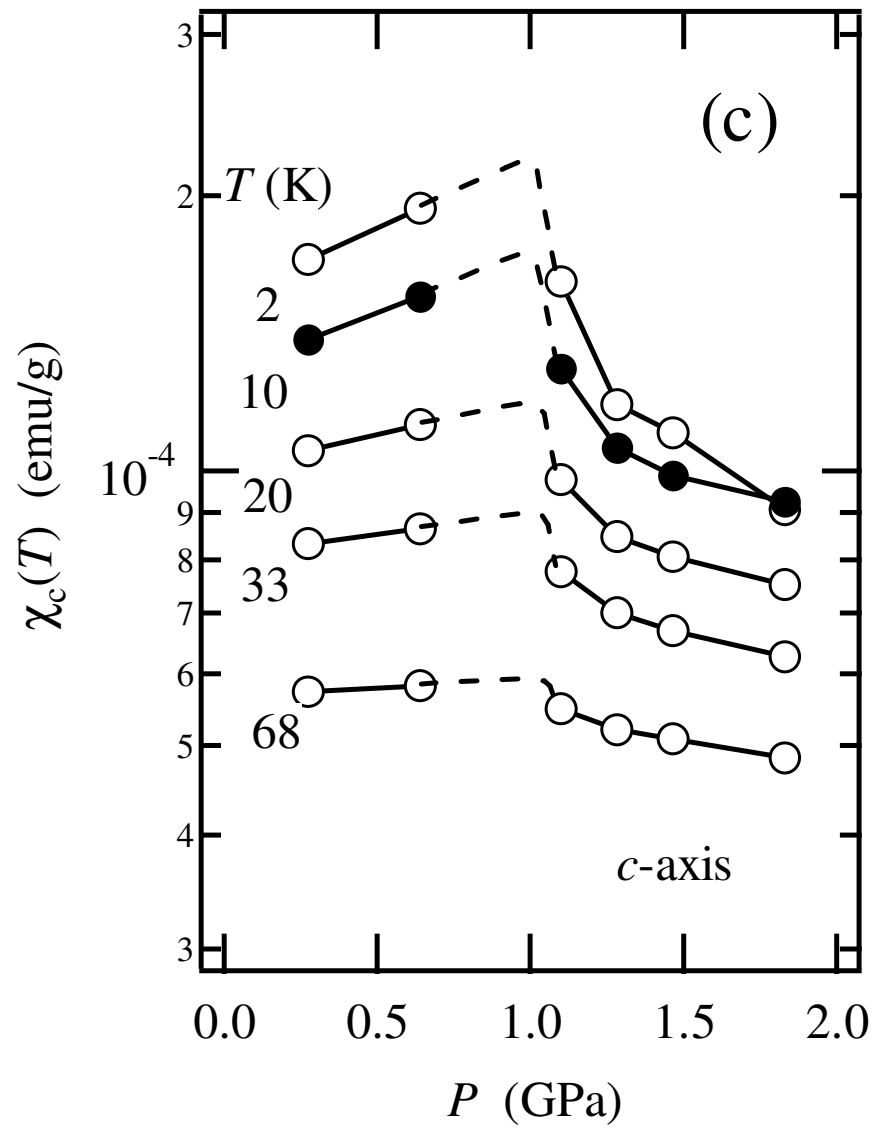

Figs. 3. (a)-(c) Pressure dependence of the magnetic susceptibility of $PrCu_2$ measured in a field of $H$ = 4 kOe directed along the *a*-, *b*- *c*-axis, respectively at various temperatures as indicated. Solid and dashed lines are guides to the eye.

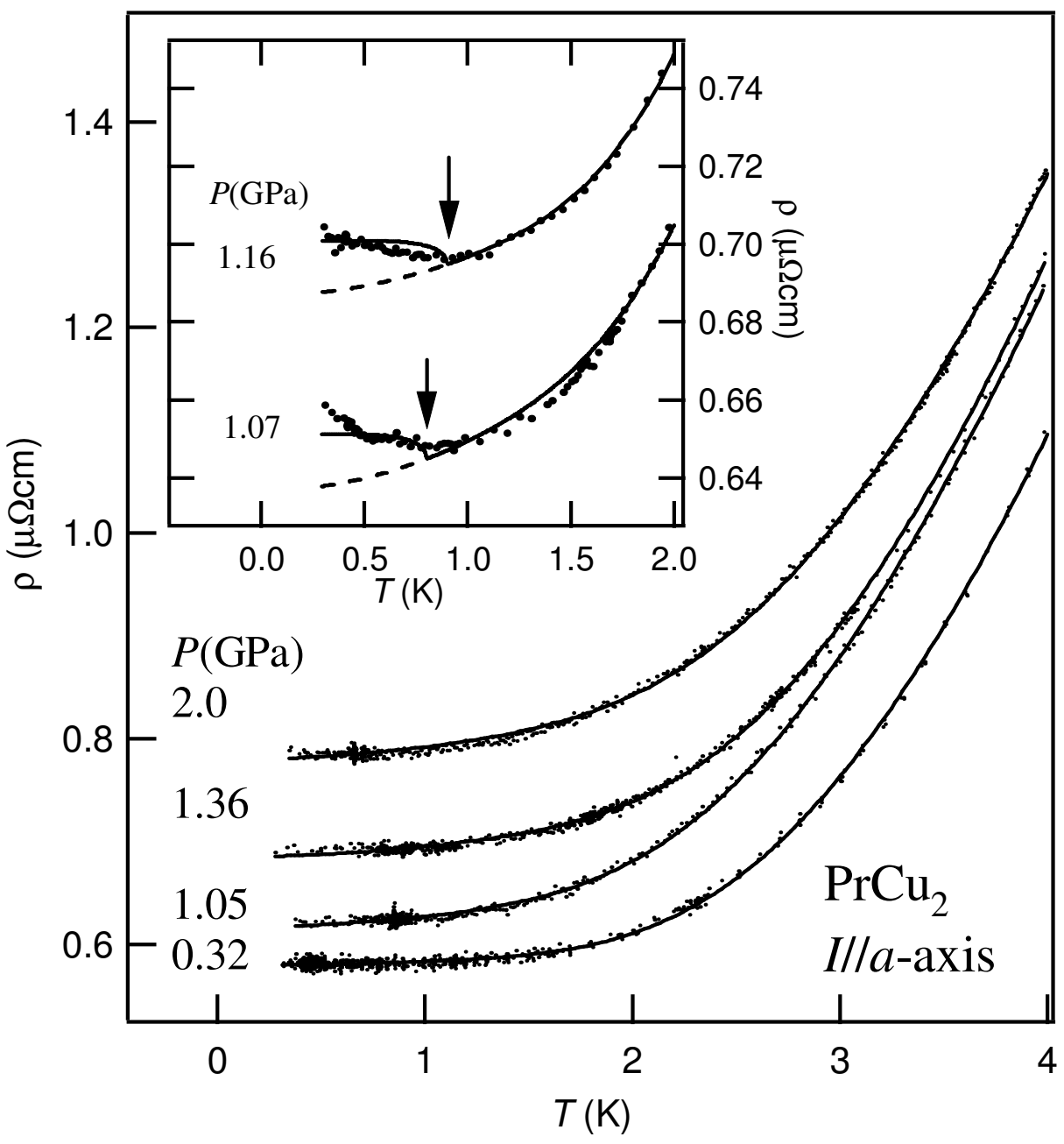

Fig. 4 Pressure variation of $\rho(T)$ at low temperatures for a current parallel to the *a*-axis. The inset shows the upturn of ρ(T) near $P=P_c$. Solid and dashed curves represent the calculated values based on eqs. (2) and (1), respectively. See text for details.

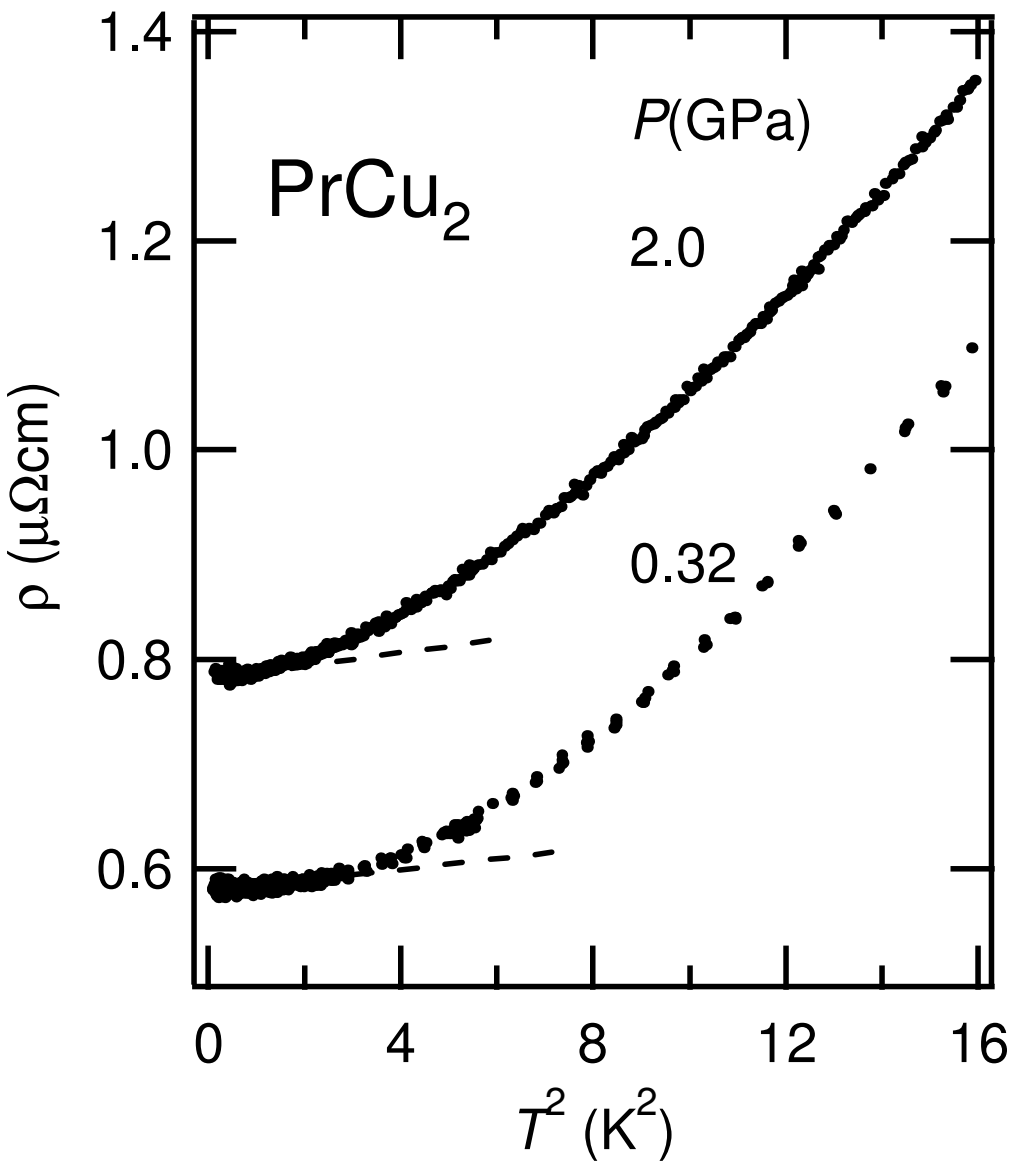

Fig. 5 Resistivity as a function of $T^2$ at various pressures. Note that $\rho(T)$ deviates clearly from a $T^2$-law (dashed lines) with increasing temperature.

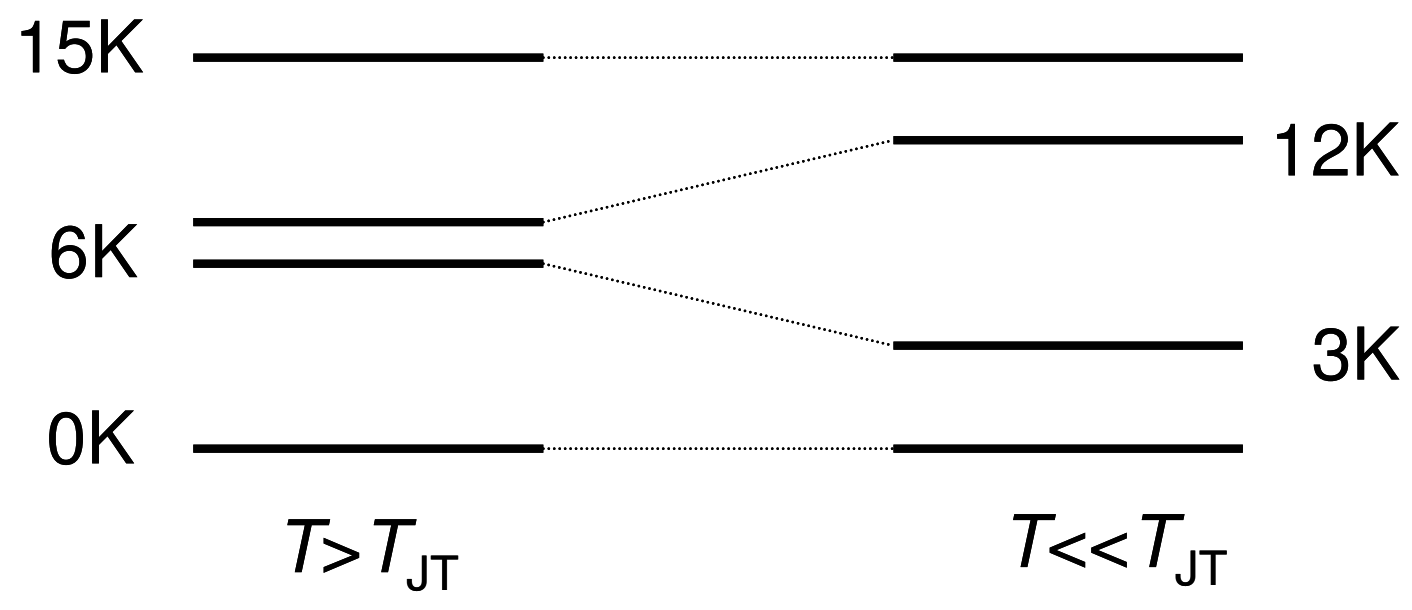

Fig. 6 The low-lying crystal field states in $PrCu_2$ at $T>T_{JT}$ and $T<<T_{JT}$ according to Ref. 12.

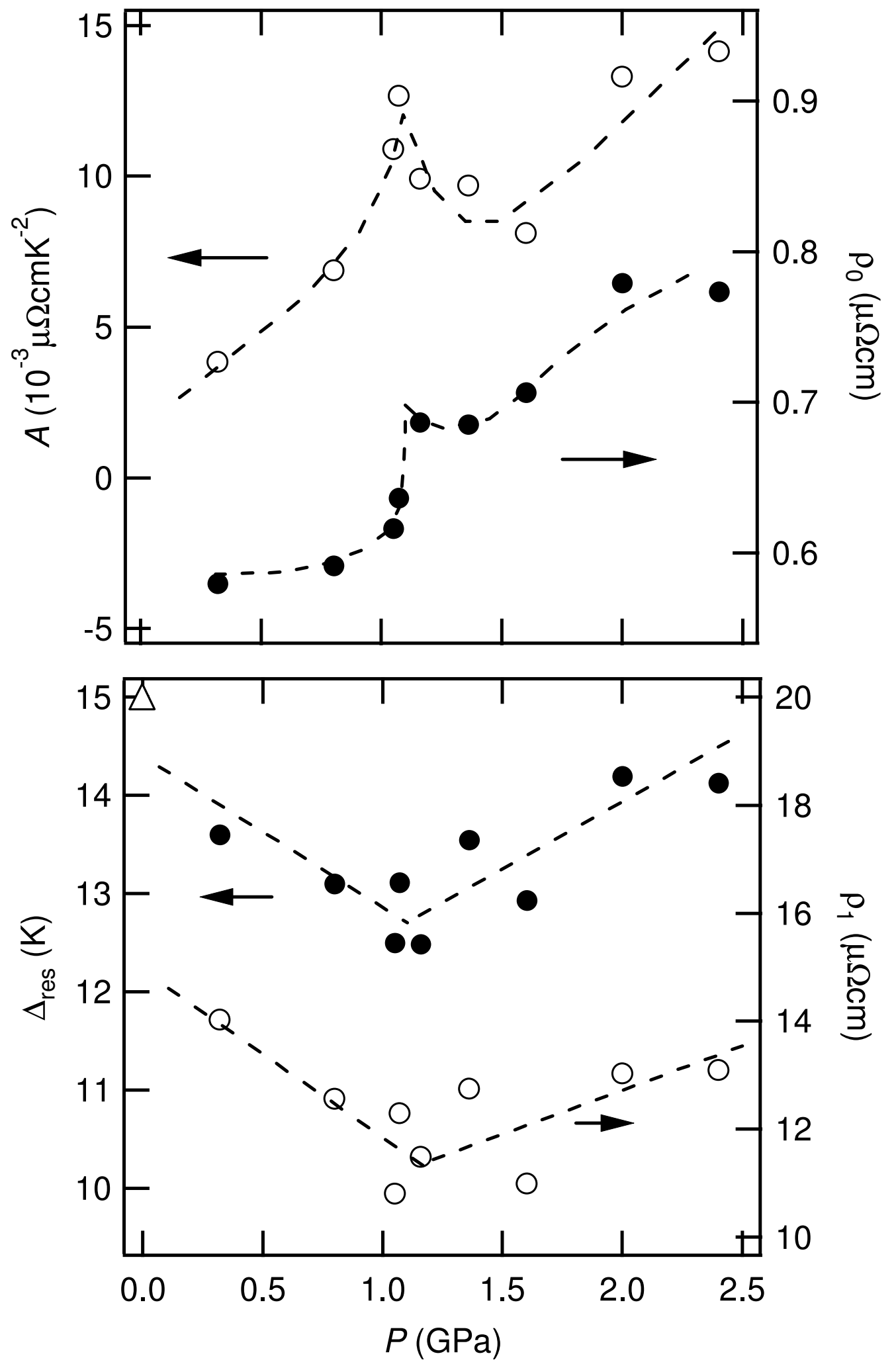

Figs. 7 Fitting parameters (a) $A$ and $\rho_0$, (b) $\Delta$ and $\rho_1$ as a function of pressure. Dashed lines are to guide the eyes. The data point (triangle) at $P$=0 is obtained from inelastic neutron scattering experiments (Ref. 12).

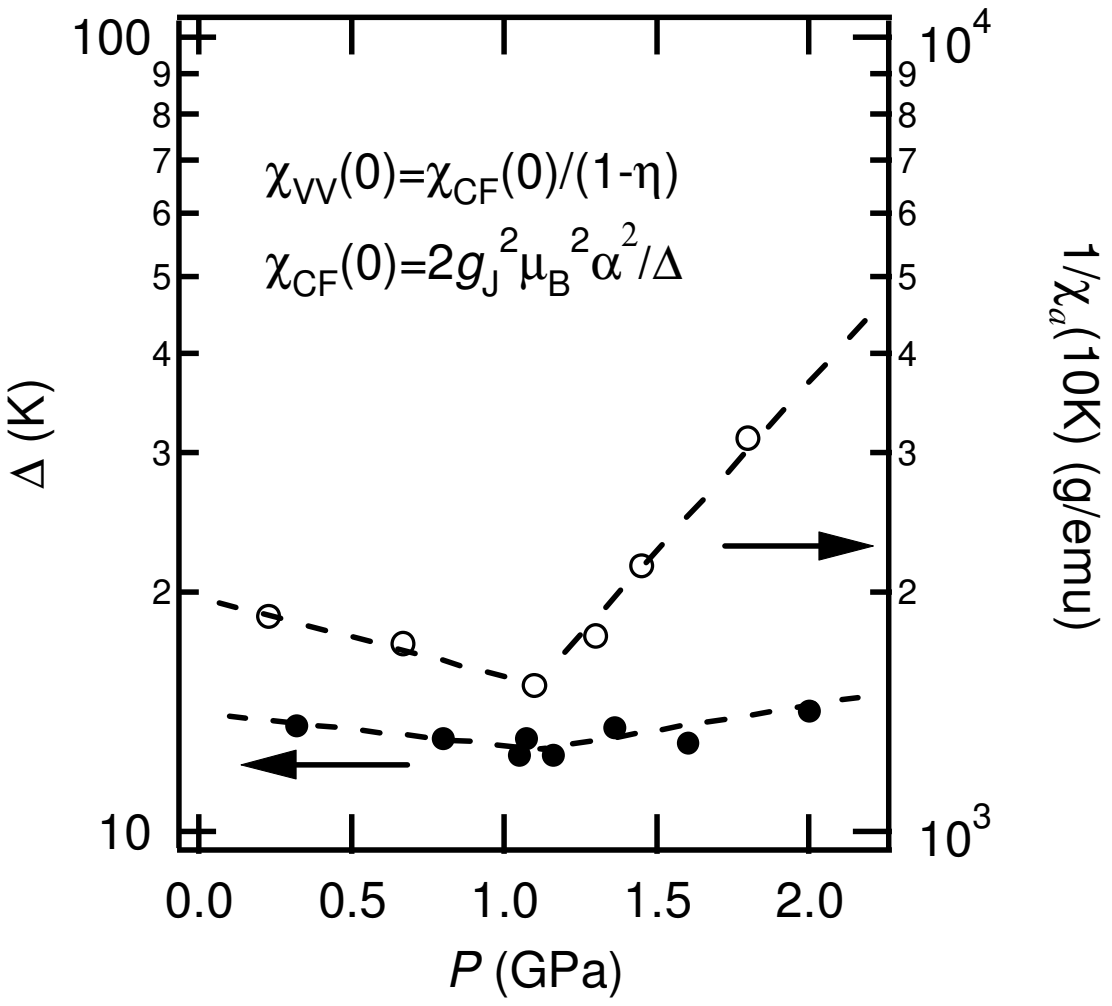

Fig. 8 $\Delta(P)$ and $\chi_a^{-1}(P)$ in a log-lin plot. Dashed lines are guides to the eye.

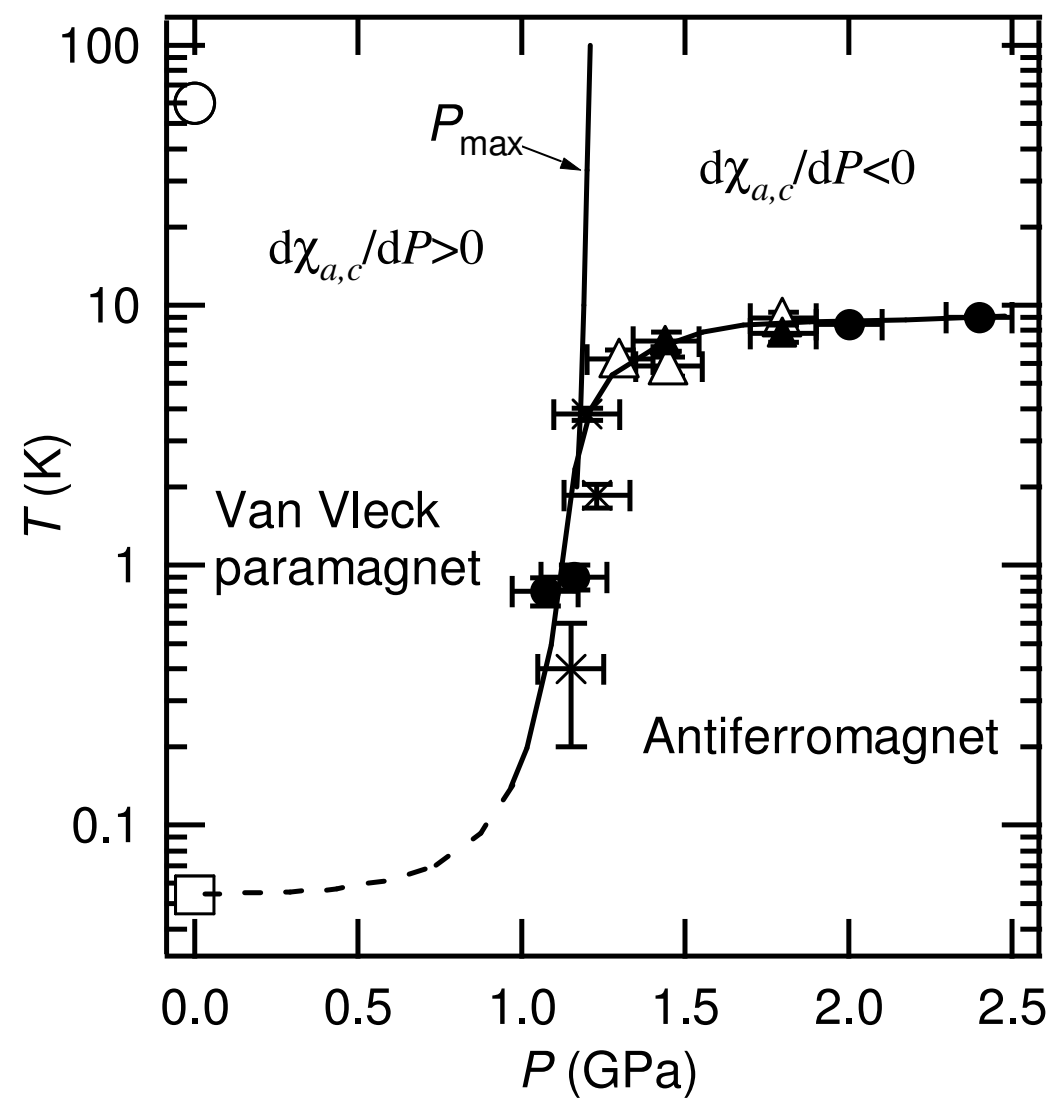

Fig. 9 Proposed magnetic phase diagram of $PrCu_2$. The ordered quadrupolar states of <$O_{22}$> seem to govern the magnetic states in the different pressure regions, respectively. The solid and dashed lines are guides to the eye. The circle and square at $P$=0 indicate the magnetic transition temperatures obtained in Refs. 9 and 7, respectively.